\documentclass{elsart} 
\usepackage{amssymb,epsfig} 

\begin{document}

\begin{frontmatter} 
 
\title{The Fadin-Lipatov-vertex in high-energy heavy quark and
charmonium production\thanksref{KSST}}
\thanks[KSST]{This talk is based on results obtained in collaboration
with R. Kirschner, A. Sch\"afer, L. Szymanowski and O.V. Teryaev.}
\author[Regensburg]{Philipp H\"agler} 
 
\address[Regensburg]{Universit\"at Regensburg, Institut f\"ur  
theoretische Physik, \newline 93040 Regensburg, Germany} 
 
\begin{abstract} 
Calculations of heavy quark and charmonium  
hadroproduction are presented in the $k_\perp$-factorization approach. 
The resulting differential cross sections are compared to experimental values 
and NLO collinear predictions. 
The application of the full effective Fadin-Lipatov-$q\bar q$-production-vertex 
has an important influence on the results and in particular leads to a  
strong suppression of certain color octet contributions which  
are related to the outstanding problem of the production of polarized $J/\Psi$. 
\end{abstract} 
 
\end{frontmatter} 
 
\section{Heavy quark production} 
 
\begin{figure}
\centerline{\epsfig{file=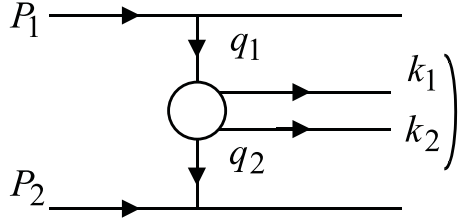,width=6cm}}
\caption{Production of a $q\bar q$-cluster in QMRK}
\label{fig1}
\end{figure}

\begin{figure}[h]
\centerline{\epsfig{file=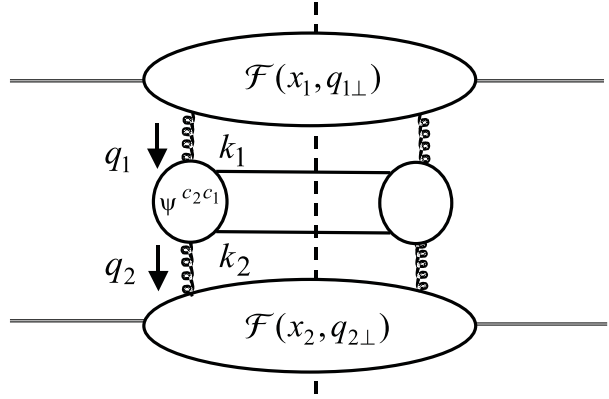,width=8cm}}
\caption{$q\bar q$-production in $k_\perp$-factorization}
\label{fig2}
\end{figure}

At very high energies $\sqrt{s}\rightarrow \infty $ in the small-$x$-region the 
Multi-Regge-Kine\-matics (MRK) respectively the Quasi-MRK 
(QMRK) is an especially suited approximation for the calculation of lepto- 
and hadroproduction cross sections. We consider the central heavy 
quark-antiquark-hadroproduction process in figs. \ref{fig1},\ref{fig2} 
and choose the 
frame where the proton momenta are given by  
\[ 
P_{1}^{+}\equiv P_{1}^{0}+P_{1}^{3}=P_{2}^{-}=\sqrt{s},\quad 
P_{1}^{-}=P_{2}^{+}=P_{1,2}^{\bot }=0. 
\] 
In the QMRK we have then an almost transverse momentum transfer $%
t_{1,2}\thickapprox q_{1,2\bot }^{2}$ and for the t-channel momenta the 
relations  
\[ 
q_{1}^{+}>>q_{2}^{+},\quad q_{2}^{-}>>q_{1}^{-}, 
\] 
but in general no strong ordering of the momenta of the produced heavy quark 
and antiquark $k_{1}\thickapprox k_{2}$. This kinematics corresponds to the 
production of a $q\bar{q}$-cluster with finite invariant mass 
and therefore makes the production of a $q%
\bar{q}$-bound state possible.
The 
associated factorization scheme is the $k_{\bot }$-factorization and the 
particles in the t-channel are allowed to have finite transverse momenta. 
The cross section for the inclusive hadroproduction of a $q\bar{q}$-pair 
at very high energies where gluon exchange dominates is then given by \cite 
{kt,Hagler00}  
\begin{eqnarray} 
\sigma _{P_{1}P_{2}\rightarrow q\overline{q}X} &=&\frac{1}{16\left( 2\pi 
\right) ^{4}}\int \frac{d^{3}k_{1}}{k_{1}^{+}}\frac{d^{3}k_{2}}{k_{2}^{+}}%
d^{2}q_{1\bot }d^{2}q_{2\bot }\delta ^{2}\left( q_{1\bot }-q_{2\bot 
}-k_{1\bot }-k_{2\bot }\right) \times   \nonumber \\ 
&&\frac{\mathcal{F}\left( x_{1},q_{1\bot }\right) }{\left( q_{1\bot 
}^{2}\right) ^{2}}\frac{\left( \psi ^{c_{2}c_{1}}\right) ^{\dagger }\psi 
^{c_{2}c_{1}}}{\left( N_{c}^{2}-1\right) ^{2}}\frac{\mathcal{F}\left( 
x_{2},q_{2\bot }\right) }{\left( q_{2\bot }^{2}\right) ^{2}}. 
\label{crosssection2} 
\end{eqnarray}
The most important parts 
of this equation are the hard scattering amplitude $\psi ^{c_{2}c_{1}}$ and 
the unintegrated gluon distribution function $\mathcal{F}\left( x_{1,2},q_{1,2\bot 
}\right) $, where the longitudinal momentum fractions are defined as 
usual, $x_{1}\equiv q_{1}^{+}/P_{1}^{+}$ and $x_{2}\equiv 
-q_{2}^{-}/P_{2}^{-}$. Due to the QMRK the gluons are off-shell and 
longitudinal, which corresponds to the replacement
$g^{\mu \nu }\thickapprox 1/(2s)P_{2}^{\mu }P_{1}^{\nu }$ 
e.g. for the numerator of the upper gluon propagator. 
The use of the 
QMRK and off-shell gluons together with the requirement of gauge invariance 
determines the $q\bar{q}$-production amplitude $\psi ^{c_{2}c_{1}}$ 
in terms of the NLLA BFKL production vertex $\Psi ^{c_{2}c_{1}}$, which is 
explicitely given by Fadin and Lipatov in \cite{Lipatov96}  
\[ 
\Psi ^{c_{2}c_{1}}\equiv \Psi ^{c_{2}c_{1}-+}=-g^{2}\left( 
T^{c_{1}}T^{c_{2}}b^{-+}(k_{1},k_{2})-T^{c_{2}}T^{c_{1}}b^{-+T}(k_{2},k_{1})%
\right) , 
\] 
where the function $b^{-+}(k_{1},k_{2})$ is defined through  
\begin{equation} 
b^{-+}(k_{1},k_{2})\equiv \gamma ^{-}\frac{~\not{q}_{1\bot }-\not{k}_{1\bot 
}-m}{\left( q_{1}-k_{1}\right) ^{2}-m^{2}}\gamma ^{+}-\frac{\gamma _{\beta 
}\Gamma ^{+-\beta }(q_{2},q_{1})}{\left( k_{1}+k_{2}\right) ^{2}},  \label{b} 
\end{equation} 
and a very similar equation holds for the crossed contribution $%
b^{-+T}(k_{2},k_{1})$. 
Equation  
(\ref{b}) includes in the second part on the right hand side
the effective gluon-production-vertex  
\begin{equation} 
\Gamma ^{+-\beta }(q_{2},q_{1})=\gamma ^{+-\beta }(q_{1},q_{2})+\left\{ 
-2t_{1}\frac{n^{-\beta }}{q_{1}^{-}-q_{2}^{-}}+2t_{2}\frac{n^{+\beta }}{%
q_{1}^{+}-q_{2}^{+}}\right\} .  \label{vertex} 
\end{equation} 
The first term of the right hand side of this equation is the standard 
three-gluon-coupling, while the second term in curly brackets is an 
additional part which is essential for gauge invariance. The function $\Gamma 
^{+-\beta }(q_{2},q_{1})$ satisfies the Ward-Slavnov-indentity $%
(q_{1}-q_{2})_{\beta }\Gamma ^{+-\beta }(q_{2},q_{1})=0$, and gauge 
invariance leads to the vanishing of the amplitude
$\psi 
^{c_{2}c_{1}}=\bar{u}(k_{1})\Psi ^{c_{2}c_{1}-+}v(k_{2})\stackrel{q_{1/2\bot 
}\rightarrow 0}{\longrightarrow }0$ in the limit of vanishing
transverse momentum. 
\begin{figure}
\centerline{\epsfig{file=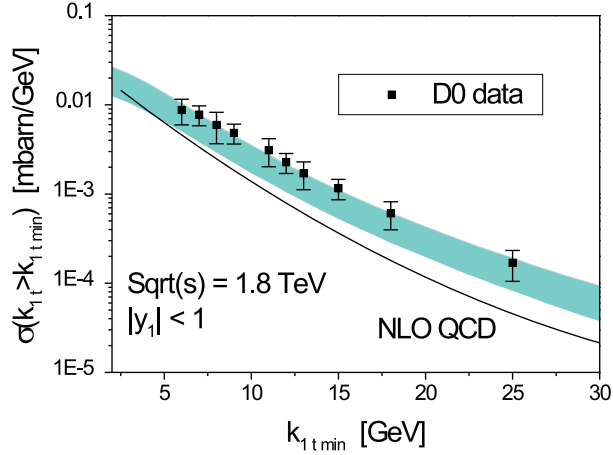,width=8cm}}
\caption{Bottom quark production}
\label{singleb}
\end{figure}

In order to be able to perform a reasonable comparison to experimental data 
we use the unintegrated gluon distribution function $\mathcal{F}\left( 
x,q_{\bot }\right) $ which has been calculated in \cite{Kwiecinski97}. The 
authors extracted $\mathcal{F}\left( x,q_{\bot }\right) $ from a fit to $%
F_{2}$ data and obtained an excellent $\chi ^{2}/$NDOF with the use of the 
following integrated gluon distribution at an input scale of $Q_{0}^{2}=1\ $%
GeV$^{2}$%
\begin{equation}
xg(x,Q_{0}^{2}=1 \mbox{GeV} ^{2})=1.57(1-x)^{2.5} 
\end{equation}
which enters the calculation through the relation  
\begin{equation} 
g(x,Q^{2})=\left( \int_{0}^{Q_{0}^{2}}+\int_{Q_{0}^{2}}^{Q^{2}}\right) 
dk_{\bot }^{2}\frac{\mathcal{F(}x,k_{\bot }^{2})}{k_{\bot }^{2}}%
=xg(x,Q_{0}^{2})+\int_{Q_{0}^{2}}^{Q^{2}}dk_{\bot }^{2}\frac{\mathcal{F(}%
x,k_{\bot }^{2})}{k_{\bot }^{2}}.  \label{integrated} 
\end{equation} 
For a 
correct treatment of the region of small transverse gluon momenta $\left| 
q_{\bot }^{2}\right| <Q_{0}^{2}$ we have to modify the cross section 
formular (\ref{crosssection2}) in a way which is compatible with equation (\ref 
{integrated}) and the vanishing of the amplitude in the limit $q_{1/2\bot 
}\rightarrow 0$, for details see \cite{Hagler00}. Our result for the 
production of a single bottom quark is shown in fig. \ref{singleb} in 
comparison to data from the Tevatron collider \cite{Abbott99} and NLO 
calculations in the collinear approach. The error band is due to a 
simultaneous variation of the bottom quark mass $4.5$ GeV $<m_{b}<4.9$ GeV 
and the fundamental scale $100$ MeV$<\lambda _{QCD}^{n_{f}=5}<180$ MeV, 
while the scale dependence of the coupling constant $\alpha _{S}(\mu ^{2})$ 
is chosen to be $\mu ^{2}=\left| q_{1,2\bot }^{2}\right| +m^{2}$. We find in 
contrast to the NLO collinear calculation a good agreement between the $%
k_{\bot }$-factorization prediction and the experimental values over the 
whole range of $k_{1\bot }$.
\begin{figure}
\centerline{\epsfig{file=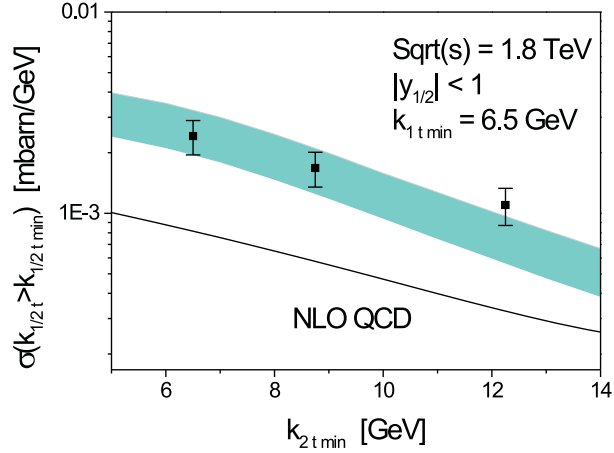,width=8cm}}
\caption{$b\bar b$-correlations}
\label{bbbar65}
\end{figure}
Likewise important 
is the comparison to measurements of $b\bar{b}$-correlations \cite{Abe97} in 
fig. \ref{bbbar65}. Allthough the cross section (\ref{crosssection2}) is only 
of LO in $\alpha _{S}$, the $k_{\bot }$-factorization together with the 
effective production vertex and off-shell gluons gives a very good 
despription of the data which lie about a factor of 3 above the NLO 
collinear result. 
 
\section{Charmonium production} 
 
\begin{figure}
\centerline{\epsfig{file=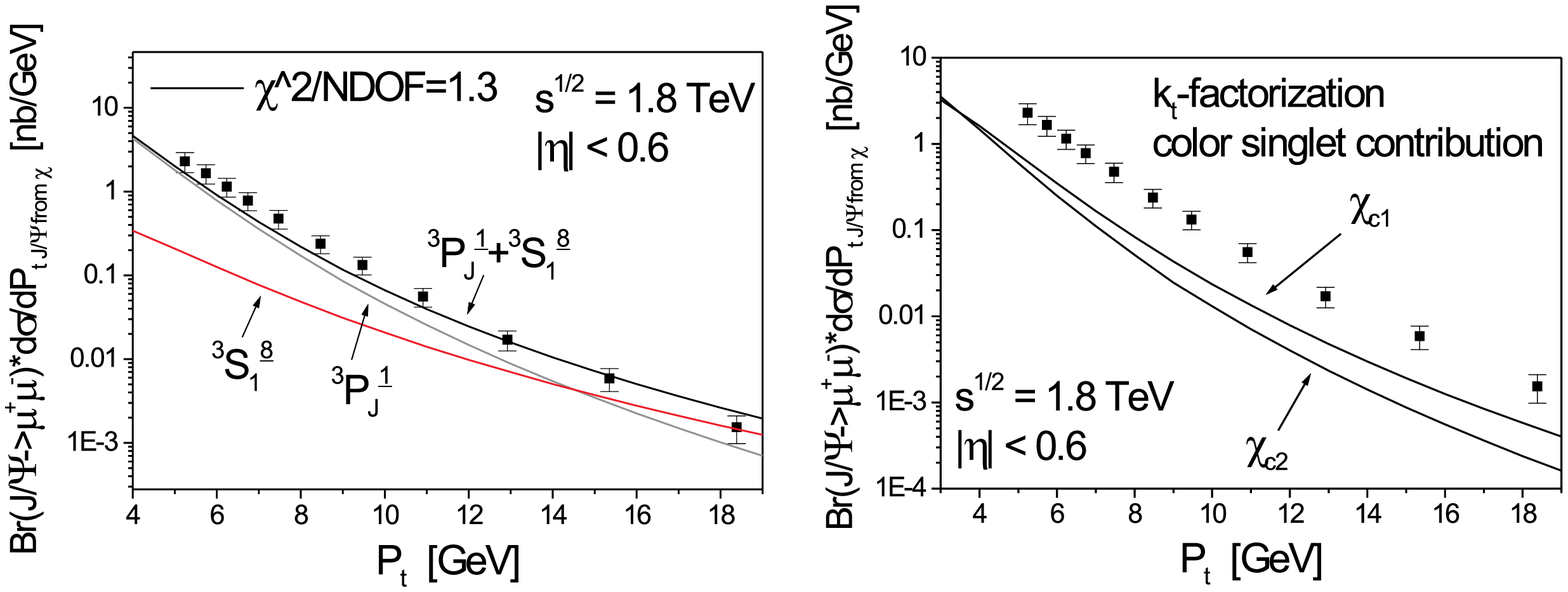,width=13.5cm}}
\caption{$\chi_c$-meson-production}
\label{chi}
\end{figure}

\begin{figure}
\centerline{\epsfig{file=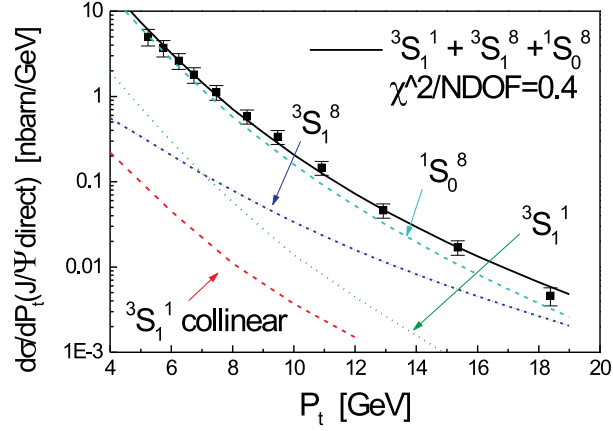,width=8cm}}
\caption{Direct $J/\Psi$-production}
\label{jpsi}
\end{figure}

Motivated by the results for heavy quark production we now pass over to 
the hadroproduction of charmonia which is a far more controversial subject  
\cite{Abe97b,Abe95}. We consider the production of the $\chi _{c}$, which is 
a $^{3}P_{J}$-state and the $J/\Psi ,$ which is a $^{3}S_{1}$ state in the 
framework of the non-relativistic-QCD (NRQCD) approach and have therefore to 
include the color singlet and color octet contributions to the final color 
singlet bound states. The calculation is again based on the cross section 
formula (\ref{crosssection2}) and the unintegrated gluon distribution from  
\cite{Kwiecinski97}. The details of the calculations and the binding 
mechanisms $c\bar{c}\rightarrow \chi _{c},J/\Psi $ are explained in \cite 
{Hagler01,Hagler02}. Because the color octet contributions depend linearly 
on the so called color octet matrix elements (COME) which at the moment 
cannot be calculated but have to be fitted to the data \cite{ChoI,ChoII}, 
the following results should not to be seen as predictions of the 
corresponding differential cross sections but are required in order to 
extract the sizes of the COME. The results for the production of the $J/\Psi  
$ from radiative $\chi _{c}$-decay in the $k_{\bot }$-factorization are 
shown on the left side of fig. \ref{chi} together with experimental data 
from the Tevatron and a NLO collinear calculation. It is clear from the 
figure that the $^{3}S_{1}^{\underline{8}}$-contribution to $\chi _{c}$ is 
generally suppressed due to the flat slope of the curve in comparison to the 
data. The flat curve originates from the additional part of the 
effective vertex in equation (\ref{vertex}) which contains factors $%
t_{1,2}\thickapprox q_{1,2\bot }^{2}$ and leads to a strong suppression of 
the corresponding COME with respect to the results from collinear 
calculations. Furthermore the Landau-Yang-Theorem, forbidding the production 
of the $\chi _{cJ=1}$-state in LO collinear factorization, is not valid in 
our approach because the gluons are off-shell. This becomes obvious in a 
rather impressive way on the right hand side of fig. \ref{chi} where
the individual $\chi _{cJ=1}$- and $\chi _{cJ=2}$-contributions are plotted. 
\newline Fig. \ref{jpsi} 
shows our results for direct $J/\Psi $-production in the $k_{\bot }$%
-factoriza\-tion framework in comparison to data and the NLO color singlet 
collinear prediction \cite{Hagler02}. Although in this case the color octet 
contributions $^{3}P_{J}^{\underline{8}}$, $^{1}S_{0}^{\underline{8}}$ 
play an important role, the $^{3}S_{1}^{\underline{8}}$-part is suppressed 
following the same reasons as in the $\chi _{c}$-production. This leads 
again to a strong reduction of the corresponding COME $\left\langle 0\right|  
\mathcal{O}^{J/\Psi }(^{3}S_{1}^{\underline{8}})\left| 0\right\rangle $ with 
respect to the collinear result. This in turn has a definite influence on 
the production of polarized $J/\Psi $ \cite{Hagler02,Yuan00} and could 
clarify the so far unsolved puzzle concerning the disagreement between 
recent experimental results and the NLO collinear predictions of transversally 
polarized charmonia \cite{Beneke97,Braaten99,Affolder00,Kniehl00}.
This work was supported by the DFG and the BMBF. 


\begin{thebibliography}{99}
\bibitem{kt} S. Catani, M. Ciafaloni and F. Hautmann, Phys. Lett.  
B242(1990) 97;
J.C. Collins and R.K. Ellis, Nucl. Phys. B360(1991) 3
\bibitem{Hagler00} Ph. H\"{a}gler, R. Kirschner, A. Sch\"{a}fer, L.  
Szymanowski, O.V. Teryaev, Phys.Rev.D 62 (2000) 071502 
\bibitem{Lipatov96}  V.S. Fadin and L.N. Lipatov, Nucl. Phys. B477(1996) 767  
\bibitem{Kwiecinski97}  J. Kwiecinski, A.D. Martin, A.M. Stasto, Phys.Rev.  
D56(1997) 3991  
\bibitem{Abbott99} B. Abbott et al., Phys.Lett B487(2000) 264 
\bibitem{Abe97} F. Abe et al., Phys.Rev. D55(1997) 2546 
\bibitem{Abe97b}  Abe et al., Phys.Rev.Lett.79(1997) 578 
\bibitem{Abe95}  Abe et al., CDF Collaboration, FERMILAB-Conf-95/226-E 
\bibitem{Hagler01} Ph. H\"{a}gler, R. Kirschner, A. Sch\"{a}fer, L.  
Szymanowski, O.V. Teryaev, Phys.Rev.Lett. 86 (2001) 1446 
\bibitem{Hagler02} Ph. H\"{a}gler, R. Kirschner, A. Sch\"{a}fer, L.  
Szymanowski, O.V. Teryaev, Phys.Rev.D63 (2001) 077501 
\bibitem{ChoI}  P. Cho, A.K. Leibovich, Phys.Rev. D53(1996) 150  
\bibitem{ChoII}  P. Cho, A.K. Leibovich, Phys.Rev. D53(1996) 6203 
\bibitem{Yuan00} F. Yuan, K.-T. Chao, Phys.Rev.Lett. 86(2001) 022002 
\bibitem{Beneke97} M. Beneke, M. Kramer, Phys.Rev. D55(1997) 5269 
\bibitem{Braaten99} E. Braaten, B.A. Kniehl, J. Lee, Phys.Rev. D62(2000) 094005 
\bibitem{Affolder00} T. Affolder et al., Phys.Rev.Lett. 85(2000) 2886 
\bibitem{Kniehl00} B.A. Kniehl, J. Lee, Phys.Rev. D62(2000) 114027 
\end{thebibliography}
\end{document}